\documentclass[10pt,conference]{IEEEtran}
\IEEEoverridecommandlockouts

\usepackage{cite}
\usepackage{amsmath,amssymb,amsfonts}
\usepackage{graphicx}
\usepackage{textcomp}
\usepackage{xcolor}
\usepackage{subfigure, booktabs}
\usepackage{adforn}
\usepackage{listings}
\usepackage{color}
\usepackage{calligra}
\usepackage{multirow}
\usepackage[normalem]{ulem}
\usepackage{indentfirst}
\usepackage{algorithm} 
\usepackage{algorithmicx}
\usepackage{algpseudocode} 
\usepackage{tcolorbox}
\usepackage{colortbl}
\usepackage{balance}
\usepackage{enumitem}

\def\BibTeX{{\rm B\kern-.05em{\sc i\kern-.025em b}\kern-.08em
    T\kern-.1667em\lower.7ex\hbox{E}\kern-.125emX}}
    
\def\name{{\sc AutoLog}}
\usepackage[]{collab}
\collabAuthor{yt}{red}{YT'Note}
\collabAuthor{yc}{blue}{YC'Note}
\collabAuthor{pj}{purple}{PJ'Note}
\collabAuthor{xzf}{magenta}{XZF'Note}
\collabAuthor{yx}{teal}{Yuxin'Note}

\definecolor{codegreen}{rgb}{0,0.6,0}
\definecolor{codegray}{rgb}{0.5,0.5,0.5}
\definecolor{codepurple}{rgb}{0.58,0,0.82}
\definecolor{backcolour}{rgb}{0.95,0.95,0.92}

\definecolor{viol}{RGB}{134,0,175}

\DeclareMathAlphabet{\mathcalligra}{T1}{calligra}{m}{n}
\usepackage{amssymb}

\begin{document}

\title{AutoLog: A Log Sequence Synthesis Framework for Anomaly Detection
}



\author{
  \IEEEauthorblockN{
    Yintong Huo\IEEEauthorrefmark{1}\IEEEauthorrefmark{2},
    Yichen Li\IEEEauthorrefmark{1}\IEEEauthorrefmark{2},
    Yuxin Su\IEEEauthorrefmark{3}\IEEEauthorrefmark{1}\IEEEauthorrefmark{1}, 
    Pinjia He\IEEEauthorrefmark{4},
    Zifan Xie\IEEEauthorrefmark{5}, and
    Michael R. Lyu\IEEEauthorrefmark{2}
  }


  \IEEEauthorblockA{\IEEEauthorrefmark{2}The Chinese University of Hong Kong, Hong Kong, China.
    Email: \{ythuo, ycli21, lyu\}@cse.cuhk.edu.hk}

  \IEEEauthorblockA{\IEEEauthorrefmark{3}Sun Yat-sen University, Zhuhai, China.
    Email: suyx35@mail.sysu.edu.cn}

  \IEEEauthorblockA{\IEEEauthorrefmark{4}Chinese University of Hong Kong, Shenzhen, China.
    Email: hepinjia@cuhk.edu.cn}
    
  \IEEEauthorblockA{\IEEEauthorrefmark{5}Huazhong University of Science and Technology, Wuhan, China.
    Email: xzff@hust.edu.cn}
}

\maketitle
\begin{abstract}
The rapid progress of modern computing systems has led to a growing interest in informative run-time logs.
Various log-based anomaly detection techniques have been proposed to ensure software reliability.
However, their implementation in the industry has been limited due to the lack of high-quality public log resources as training datasets.

While some log datasets are available for anomaly detection, they suffer from limitations in (1) comprehensiveness of log events; (2) scalability over diverse systems; and (3) flexibility of log utility.
To address these limitations, we propose \name, the first automated log generation methodology for anomaly detection.
\name~uses program analysis to generate run-time log sequences without actually running the system. 
\name~starts with probing comprehensive logging statements associated with the call graphs of an application. Then, it constructs execution graphs for each method after pruning the call graphs to find log-related execution paths in a scalable manner. 
Finally, \name~propagates the anomaly label to each acquired execution path based on human knowledge.
It generates flexible log sequences by walking along the log execution paths with controllable parameters.
Experiments on 50 popular Java projects show that \name~acquires significantly more (\textit{9x-58x}) log events than existing log datasets from the same system, and generates log messages much faster (\textit{15x}) with a single machine than existing passive data collection approaches.
\name~also provides hyper-parameters to adjust the data size, anomaly rate, and component indicator for simulating different real-world scenarios.
We further demonstrate \name's practicality by showing that \name~enables log-based anomaly detectors to achieve better performance (\textit{1.93\%}) compared to existing log datasets.
We hope \name~can facilitate the benchmarking and adoption of automated log analysis techniques.\let\thefootnote\relax\footnotetext{$^*$Co-first authors.}\footnotetext{$^{**}$Corresponding author.}





\end{abstract}

\maketitle
\section{Introduction}

The ever-growing amount and quality of data cultivate the development of advanced data-driven approaches, showing great power in many fields. Considering the dataset evolving from the simple handwritten digit dataset MNIST~\cite{lecun1998gradient} to the large visual database ImageNet~\cite{deng2009imagenet}, the rich data foster more sophisticated and applicable image algorithms against the complicated real-world scenarios.


The log analysis community has proposed numerous techniques to assist maintainers in automatically inspecting the large log files produced by modern complex systems.
The techniques include detecting internal system anomalies~\cite{li2020swisslog, yang2021plelog, du2017deeplog, huo2021semparser, liu2023scalable}, and suggesting the root cause of an anomaly~\cite{wang2021groot, lu2017log, wang2020root,path,li2022intelligent} for employing the corresponding mitigation strategy. 
Despite the continued development of automated solutions for software maintenance through log analysis, there has been a lack of emphasis on the need for effective datasets.
Consequently, only a few of these techniques have been successfully deployed in real-world settings.
This highlights the gap between the limited log data used in academic research and the complexity of industry deployment~\cite{zhao2021empirical}.
 \begin{figure}[tbp]
     \centering
     \includegraphics[width=\columnwidth]{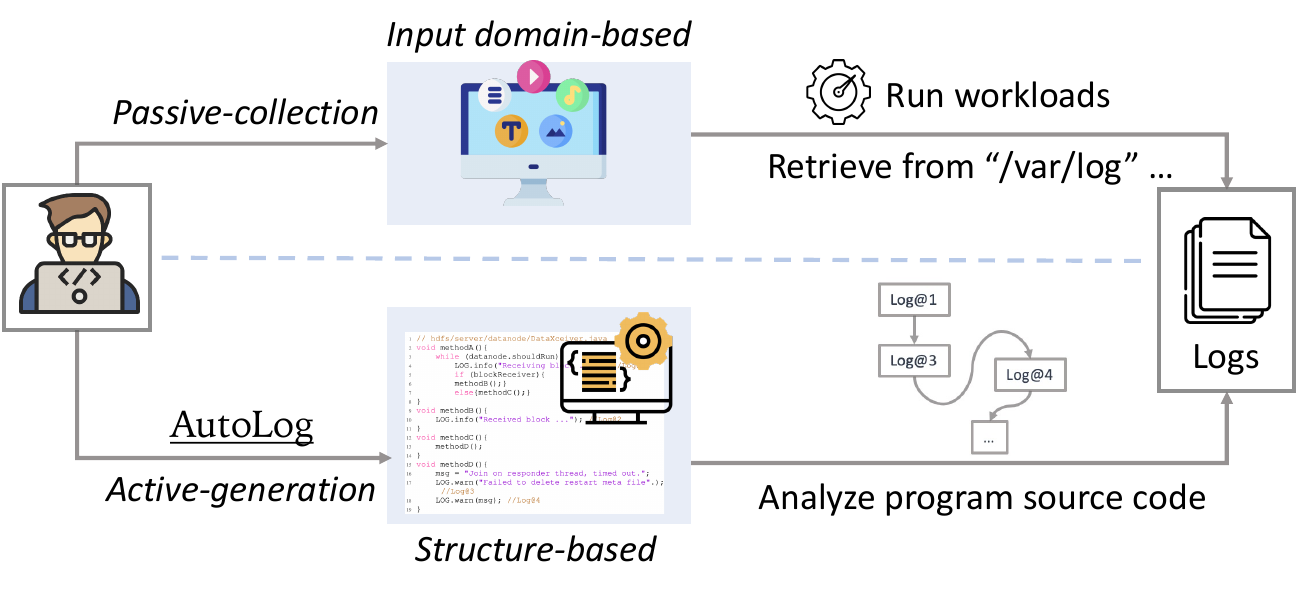}
\vspace{-0.15in}
     \caption{Difference of existing passive-collection approach and our active-generation methodology \name.}
     \label{fig:intro-difference}

 \end{figure}



Acquiring sufficient, high-quality, and representative logs for practical analysis is challenging.
On the one hand, industrial logs from real-world large service providers~\cite{zhao2021empirical, jia2017approach, zhang2021onion, jia2022augmenting} (e.g., IBM, Microsoft) contain rich events but have privacy concerns, making it difficult to release them publicly for researchers.
On the other hand, logs collected in laboratory environments using simulation~\cite{lin2016log, xu2009detecting, he2020loghub} are publicly available but contain simple events produced from limited standard workloads.
As a result, they do not accurately reflects the complex run-time workflow of large-scale software.
In either case, existing datasets are based on a \textit{passive-collection methodology} that retrieves log files after running applications in a computing system, as described in Fig.~\ref{fig:intro-difference} (Up).
The output logs from the passive-collection methodology are solely dependent on the value of inputs, which impedes exploring and evaluating models on three aspects, namely:




(1) \textit{Comprehensiveness} of log events. 
Log events are formulated as logging statements in source code for execution.
When collecting logs passively, the diversity of log events is mainly determined by the variety of input workloads. 
For example, the widely-used Hadoop dataset~\cite{lin2016log} was generated using only two test applications.
Even though the system is deployed in scale, it is still impractical to traverse all \textit{what-if} scenarios~\cite{wang2021tsagen} during execution to trigger complete logging statements by deliberately designed workloads. 



(2) \textit{Scalability} over diverse systems. 
Collecting logs from new systems requires engineers to re-deploy the system and re-implement workloads, which is time-consuming and labor-intensive~\cite{he2020loghub, oliner2007supercomputers}.
Hence, existing datasets are inadequate in terms of system logs, making it challenging to evaluate the generalization ability of proposed log analysis techniques across diverse systems. For example, the most popular dataset, LogHub~\cite{he2020loghub}, integrates only five labeled system logs for evaluating anomaly detectors. 



(3) \textit{Flexibility} of log utility. 
While existing datasets allow for minor modifications (e.g., anomaly rate) for evaluation purposes~\cite{chen2021experience, le2022log}, these changes are inadequate for imitating diversified scenarios.
For example, when maintainers need to analyze the functionality of a storage component, existing datasets with anomaly-or-not annotation cannot distinguish these component-specific logs.
Hence, the complicated conditions in the actual production environment necessitate a more flexible and controllable log collection methodology. 




The limitations of the current passive-collection methodology have led to the need for effective log collection for building practical solutions. 
Since system logs are generated during the execution of logging statements in the source code, (e.g., \texttt{LOG.error(``failed to start web server.'')}),
we consider the log collection problem as the task of constructing log sequences based on the execution order of logging statements. 
To create log sequences, we adopt the idea from program analysis to develop~\name, the first \uline{Auto}mated \uline{Log} generation methodology that \textit{actively generates} effective datasets for anomaly detection (Fig.~\ref{fig:intro-difference} (Down)). The novel static-guided approach can uncover execution paths and produce rational log sequences~\cite{chen2018automated,yuan2012improving, zhao2017log20}.

Specifically, \name~generates effective log datasets with three phases: 
(1) \textit{logging statement probing phase} 
explores all methods containing the logging statements to achieve \textit{comprehensive} log events coverage. 
(2) \textit{log-related execution path finding phase}, which prunes the call graphs and acquires the execution paths related to the logging statements to ensure \textit{scalability}.
(3) \textit{log graph walking phase}, which forms log sequences from execution paths and labels the anomaly sequences based on expert annotations. The controllable parameters of \name~enable researchers to customize the log dataset with different data sizes, anomaly ratios, and component indicators, providing \textit{flexibility} in log utility.

We conduct an extensive evaluation of \name~to 50 popular Java projects and compare the resulting dataset to existing passively-collected datasets,
showing its superiority from three perspectives:
(1) \name~generates \texttt{9x-58x} more log events than existing datasets from the same system, and covers \textit{87.77\%} of all log events of the 50 projects on average.
(2) \name~successfully produces logs at least \texttt{15x} faster than the collection time for existing log datasets, with tens of thousands of messages per minute on average.
(3) \name~is built with options to change the data size, anomaly ratio, and certain components of logs, supporting a wide range of development environment simulations.
Further, we show that existing anomaly detection models effectively gain improvements (1.93\%) by learning from the comprehensive log dataset generated by \name.
Thus, we believe that \textit{the data generation methodology \name~can serve as a critical resource for developing advanced log analysis algorithms, as well as for providing testing and benchmarking data for such algorithms to ensure software reliability.} 

To summarize, the contributions of this paper are:
\begin{itemize}
    \item We present \name, a novel, widely-applicable automated log generation methodology that addresses three limitations of existing log datasets: lack of comprehensiveness, scalability, and flexibility.
    \item \name~has three phases: logging statement probing, log-related execution path finding, and log path walking.
    \item Extensive experiments show that \name~achieves a comprehensive (\textit{87.77\%}) coverage of log events and efficiently produces log datasets (over 10,000 messages/min) on Java projects, and offers the flexibility for adapting to multiple scenarios. We further demonstrate that \name~improves anomaly detectors by 1.93\%.
    \item To our knowledge, \name~is the first log sequence generator. All artifacts and datasets are released for future research.\footnote{https://github.com/logpai/AutoLog.}

\end{itemize}



\lstset{
    commentstyle=\color{brown},
    keywordstyle=\color{magenta},
    numberstyle=\tiny\color{codegray},
    stringstyle=\color{codepurple},
    basicstyle=\ttfamily\footnotesize,
    breakatwhitespace=false,         
    breaklines=true,                 
    captionpos=b,                    
    keepspaces=true,                 
    numbers=left,                    
    numbersep=5pt,                  
    showspaces=false,                
    showstringspaces=false,
    showtabs=false,                  
    tabsize=2,
    language=Java
}

\begin{lstlisting}[caption=A simplified example from HDFS., label={lst:emp-case}]
// hdfs/server/datanode/DataXceiver.java
void methodA(){
    while (datanode.shouldRun){
        LOG.info("Receiving block " + block); //Log@1
        if (blockReceiver){
            methodB();}
        else{
            methodC();}
    }
}        
void methodB(){
    LOG.info("Received block " + block); //Log@2
} 
void methodC(){
    methodD();
}
void methodD(){
    msg = "Join on responder thread, timed out.";
    LOG.warn("Failed to delete restart meta file."); //Log@3
    LOG.warn(msg); //Log@4
} 

\end{lstlisting}

\section{Motivating study}\label{sec:empirical}
\begin{table*}[tbp]
\small
\centering
\caption{Statistics and descriptions of existing datasets. \# Log Event, \# Workload, \# Failure Type, and \# Message show the number of log events, executed workloads, failure types, and log messages in each dataset, respectively.}

\label{tab:emp}
\begin{tabular}{c||c|c|c|c|c|c}
\toprule
\textbf{Dataset}   & \textbf{\# Log Event} & \textbf{\# Workload} & \textbf{\# Failure Type} & \textbf{\# Message} & \textbf{Collection Time} & \textbf{Annotation}\\ \toprule
$\mathcal{D}$-HDFS  & 30 & NA & 11 & 11,175,629 & 38.7 Hours & With labelling\\
$\mathcal{D}$-Hadoop & 242 & 2 & 3 & 394,308 & NA & With labelling \\
$\mathcal{D}$-BGL & 619 & NA & NA & 4,747,963 & 214.7 days & With labelling\\
$\mathcal{D}$-Zookeeper & 77 & NA & NA & 207,820 & 26.7 days & Without labelling\\
\bottomrule
\end{tabular}
\end{table*}

\subsection{Study Subject}
We select four widely-used datasets (Table~\ref{tab:emp}) as subjects released by different project teams, including distributed systems and supercomputers.
In particular, $\mathcal{D}$-HDFS~\cite{xu2009detecting}, $\mathcal{D}$-Hadoop~\cite{lin2016log}, and $\mathcal{D}$-BGL~\cite{oliner2007supercomputers} datasets are collected for anomaly detection after running normal workloads and injecting several types of failures in each system, respectively. $\mathcal{D}$-Zookeeper~\cite{he2020loghub} is collected by running several benchmarks without labeling.

\subsection{Are Existing Datasets Comprehensive?}

We first investigate whether the existing log datasets provide comprehensive coverage of logging statements.
Considering the case in Listing~\ref{lst:emp-case} from Hadoop, \texttt{Log@2} will occur if $\mathtt{blockReceiver}$ is enabled;
otherwise, \texttt{Log@3} and \texttt{Log@4} will be logged.
Nevertheless, we find that the latter were absent in the $\mathcal{D}$-HDFS dataset, likely due to $\mathtt{blockReceiver}$ being enabled at all times.
As a result, the anomaly detection techniques trained with an inadequate dataset may fail to tackle the unseen log, and even their experiment conclusions may not be representative.
For instance, neglecting the responder connection time anomaly, which is associated with \texttt{Log@3} and \texttt{Log@4}, can happen if the dataset used for training lacks these logs.
Furthermore, Table~\ref{tab:emp} reveals that the existing datasets collected through passive-collection approaches are limited in the number of failure types and workloads they cover.
The literature~\cite{cinque2010assessing,pecchia2012detection} implies that although a range of workloads has been implemented, it is still unrealistic to inject all types of anomalies to trigger all log events. Thus, these datasets fall short of comprehensive coverage, creating a significant gap with real-world scenarios.


\subsection{Are Existing Datasets Scalable?}

Validating the generalization ability over diverse systems is critical for practical log analysis algorithms.
We use the term scalability to measure the effort (e.g., time, workforce) we should put in to acquire logs from multiple system sources.

To determine the scalability of existing datasets, we summarize their log message amounts and collection time in Table~\ref{tab:emp}. The table reveals that 
it takes a long time to collect logs, even after the system has been deployed and configured. For example, collecting 207,820 log messages from the Zookeeper takes
more than 26 days. 
Additionally, since existing datasets are obtained from running system applications, it requires additional expert efforts to redeploy and rerun the applications when extending the workloads to a new system. 
Unfortunately, software maintainers cannot afford to wait such a long time before developing or validating new algorithms. In short, existing datasets are not scalable enough, necessitating the exploration of a new, more efficient approach to log collection.




\subsection{Are Existing Datasets Flexible?}
Previous research has examined the impact of certain dataset characteristics, such as data distribution and data selection, on log-based anomaly detectors~\cite{chen2021experience, le2022log}.
However, to further investigate the performance of log analytics, it is necessary to customize log datasets to simulate diverse scenarios, which requires flexibility.

Ideally, all data can be collected, but in reality, data cannot be completely collected (or collected in a large enough quantity), and they suffer from restrictive flexibility.
For example, to increase the anomaly ratio from 0.1\% to 15\%, researchers may need to remove a large number of normal logs (\texttt{150x}) from the existing dataset, sacrificing data quantity~\cite{le2022log}.
Moreover, logs collected from system files without component specifications raise difficulties for component-wise analysis. More ingredients of the flexibility are elaborated in Section~\ref{sec:RQ3}.
In this regard, existing datasets cannot flexibly demonstrate model effectiveness under various scenarios, motivating a controllable log sequence acquisition approach.




\section{Methodology}\label{sec:methodology}

\subsection{Overview}
Log files are created every time when a system executes logging statements. Therefore, we view the log sequence generation problem as a task of finding the execution paths related to logging statements in the program.
Generating log sequences for anomaly detection involves two primary questions: how to generate program execution paths that include logging statements (Phase2), and how to identify whether the execution path produces an anomaly or not (Phase3).

\begin{figure*}[htbp]
    \centering
    \includegraphics[width=\linewidth]{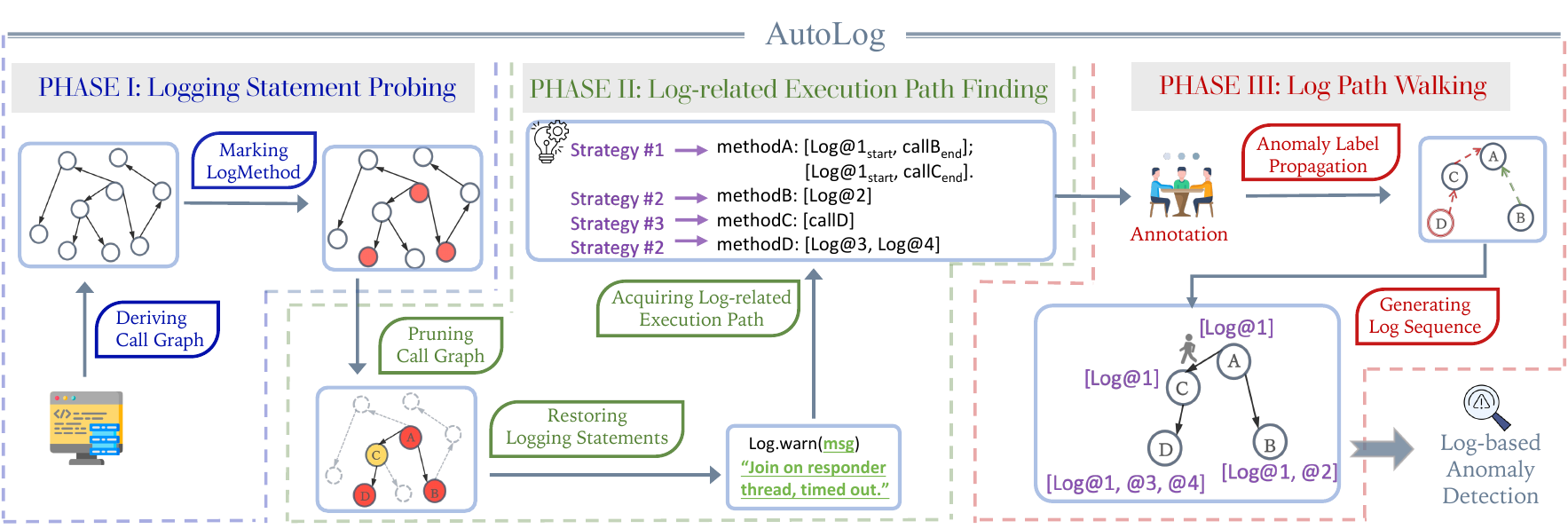}
    \vspace{-0.1in}
    \caption{The overall framework of \name~with three phases: logging statement probing, log-related execution path finding, and log path walking. The details of "Acquiring Log-related Execution Path" are illustrated in Fig.~\ref{fig:path}.}
    \label{fig:autolog-framework}
\end{figure*}

To tackle the problem, \name~takes the program as input and outputs a dataset for anomaly detection by analyzing its execution paths.
It comprises three phases, as illustrated in Fig.~\ref{fig:autolog-framework}:
The first phase builds call graphs for the project and marks all methods that contain logging statements (LogMethod), enabling \textit{comprehensive} coverage of log events. 
The second phase prunes out the call graph nodes and records the log-related execution paths (LogEPs) within each remaining method to overcome the \textit{scalability} issue. 
The third phase propagates logging statement labels from domain experts to all LogEPs. \name~eventually generates \textit{flexible} log sequences with chaining LogEPs across methods based on their calling relationship.

\subsection{PHASE1: Logging Statement Probing}
To cover comprehensive log events, \name~starts with exploring the methods containing logging statements and the calling relationships of these methods.

\subsubsection{Deriving Call Graphs}
Initially, we perform a standard static analysis to construct a call graph that chains methods based on their execution-time relationships in a program~\cite{ryder1979constructing}.
A call graph is a directed graph where each node identifies a method and each edge represents a pair of (caller, callee) method-level relationships. 
\name~derives call graphs with the context-insensitive pointer analysis~\cite{smaragdakis2015pointer} to enhance the precision of the call graph in handling virtual method calls, calls through interfaces, and polymorphism.
Afterwards, we treat each call graph as a directed acyclic graph after marking the cycles induced by recursion invocation. 



\subsubsection{Marking LogMethod}
Since we concentrate on logs, \name~identifies methods containing logging statements (LogMethods) by checking whether any logging API is used in a method.
To detect a variety of logging APIs used in different projects, we summarize commonly-used logging frameworks (e.g., slf4j~\cite{slf4j}) and capture logging APIs by analyzing the invocation of these popular frameworks. 
Compared to the regular expressions applied in LogCoCo~\cite{chen2018automated}, this API-based analysis is capable of recognizing more comprehensive customized logging APIs by examining the inheritance relationships.
Fig.~\ref{fig:autolog-framework} illustrates how \name~manages Listing~\ref{lst:emp-case}, where $\mathtt{methodA}$, $\mathtt{methodB}$, and $\mathtt{methodD}$ are recognized as LogMethod (highlighted in red).




\subsection{PHASE2: Log-related Execution Path Finding}
An intuitive idea to solve the execution path finding problem is to construct the entire execution graph of a project and traverse it.
However, in most cases, large-scale software contains an infinite number of paths~\cite{taylor1992structural} so that exhaustive enumeration undoubtedly causes path explosion problem~\cite{nejmeh1988npath}.
To overcome scalability challenges, \name~takes two steps: (1) pruning out the call graph nodes that will not induce LogMethods; (2) traversing the intra-method execution graph and recording the execution paths that are related only to logging activities. 

\subsubsection{Pruning Call Graph}
The goal of pruning is to eliminate redundant nodes from the call graph, particularly those that neither represent LogMethod nodes nor nodes that lead to any LogMethod nodes.
A node may induce a LogMethod node in two ways: (1) by calling it directly, or (2) by calling it indirectly through other intermediate nodes.
Identifying the LogMethod-inducing nodes can be viewed as a graph sorting problem that finds all ancestor nodes of specific nodes (i.e., LogMethod nodes) in the graph. To do so, \name~performs topological sorting~\cite{graham1982gprof} over the call graph.
Using the topological sorting, a node is considered a non-LogMethod-inducing method if it is neither a LogMethod nor comes before any LogMethod, indicating that it is not an ancestor of any LogMethod. 
In Fig.~\ref{fig:autolog-framework}, red, yellow, and dashed contour nodes are used to represent LogMethod, LogMethod-inducing methods, and non-LogMethod-inducing methods, respectively.

All non-LogMethod-inducing nodes and the edges associated with them (represented by dashed lines) will be pruned out.
The resulting pruned graph (denoted as $CG'$) is much smaller than the original call graph since many methods do not contain or induce other logging statements.
For example, only \textit{14.79\%} of nodes are reserved after pruning call graphs for the HDFS system.
Because only the remaining methods are used for further analysis, the pruning step significantly promotes \name's efficiency and facilitates subsequent analysis.

 \begin{figure}[t]
     \centering
     \includegraphics[width=\columnwidth]{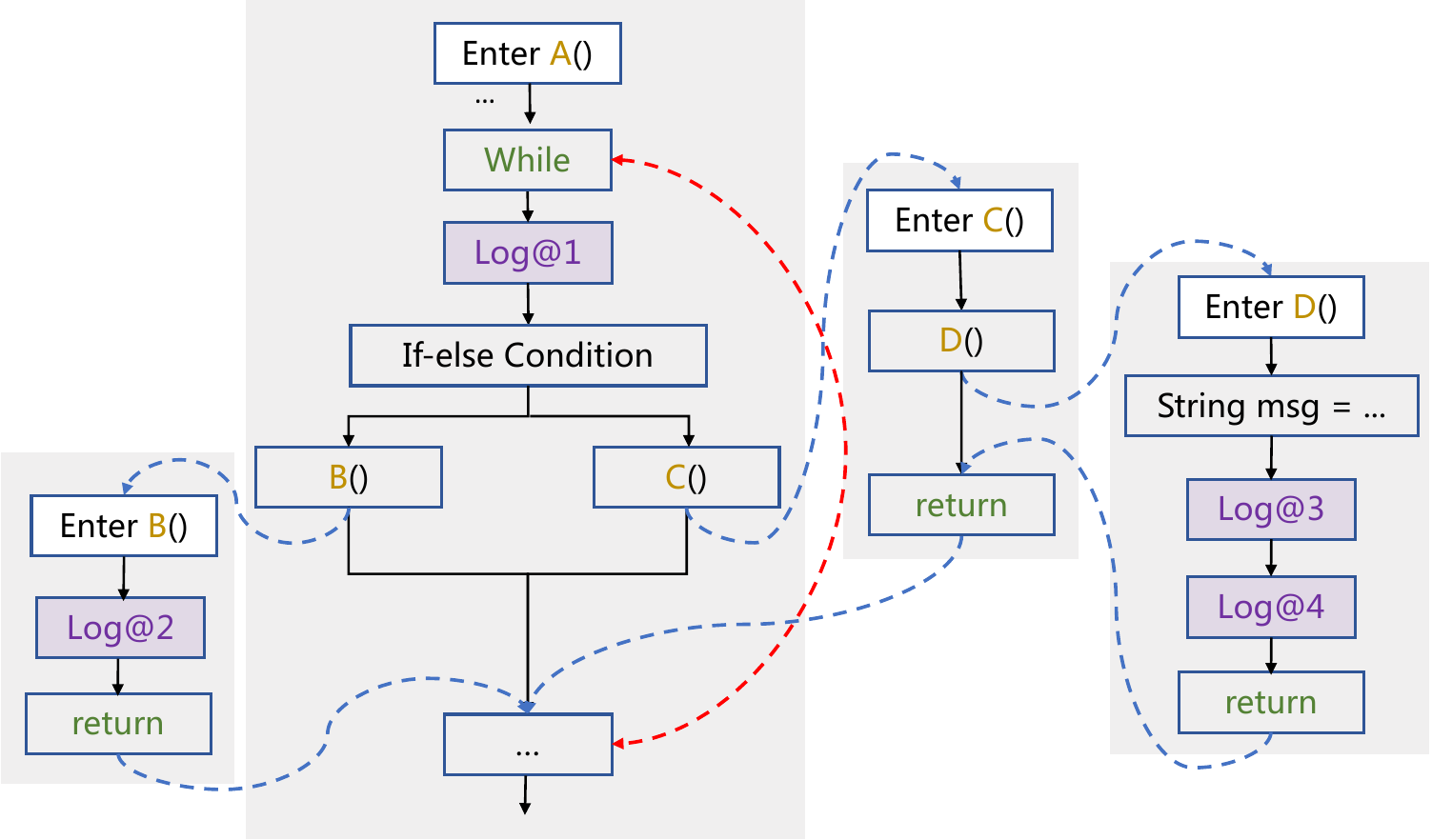}
     \vspace{-0.1in}
     \caption{The simplified execution graphs for methods in Listing~\ref{lst:emp-case}.}
     \label{fig:path}
    \vspace{-0.1in}
 \end{figure}

\subsubsection{Restoring Logging Statements}
This step involves restoring logging statements by resolving run-time parameters to provide more detailed information beyond the specific logging statements.
A logging statement typically includes constant strings written by developers (e.g., ``\texttt{Receiving block}'') and run-time parameters (e.g., $\mathtt{block}$). 
Specifically, \name~resolves the parameters that are constant string variables inside the methods. For example, the parameter $\mathtt{msg}$ (Line20) is resolved from its assignment (Line18) in Listing.~\ref{lst:emp-case}.
For other types of parameters (e.g., $\mathtt{block}$), \name~replaces them with a dummy token (i.e., $<$*$>$) since they are typically removed during preprocessing for anomaly detection models~\cite{zhang2019robust, lu2018detecting}. 

\subsubsection{Acquiring Log-related Execution Paths}
Log sequences generate along with software run-time, reflecting the control-flow paths and execution states~\cite{yuan2012improving, agrawal1998mining,amar2018using,busany2016behavioral}.
However, obtaining the order of realistic log events by enumerating execution paths from the entire application can be challenging due to their huge size~\cite{kuznetsov2012efficient}.
To solve this issue, \name~constructs and traverses the small-scale execution graph for each method in the $CG'$ to enable scalability.

Particularly, \name~builds an execution graph with control flow information for each method and links the invocation with the corresponding method in the entry and exit points. Each node in the execution graph represents an executed activity, and the edge represents the relationship between two activities in the temporal order.


Afterwards, \name~traverses each execution graph and only records log-related execution paths (\textit{LogEPs}) for each method, which includes the \textit{invocations} and \textit{logging activities} (i.e., logging statements). The invocation is noteworthy because a log sequence in a method can be interrupted by another log sequence introduced by the invocation. \name~derives LogEPs for each method by using three strategies. Fig.~\ref{fig:path} depicts the execution graphs for four methods of Listing~\ref{lst:emp-case} to illustrate the strategies.



\begin{itemize}

\item[\adfhalfrightarrowhead] Strategy \#1: For nodes that are \textit{both non-leaf nodes and LogMethod} in the $CG'$, \name~considers the execution order of invocations and logging activities. In our case, we obtain two LogEPs for $\mathtt{methodA}$: \texttt{[Log@1, callB]}, and \texttt{[Log@1, callC]}.

\item[\adfhalfrightarrowhead] Strategy \#2: For \textit{leaf nodes} in $CG'$ that are definitely \textit{LogMethods}, \name~enumerates all possible execution sequence of logging activities. This strategy is applied to $\mathtt{methodB}$ and $\mathtt{methodD}$, leading to the LogEP of \texttt{[Log@2]} and \texttt{[Log@3, Log@4]}, respectively.

\item[\adfhalfrightarrowhead] Strategy \#3: For the \textit{non-leaf and non-LogMethod} nodes in the $CG'$, \name~records all possible invocation sequences in execution.
In this case, one LogEP is derived for $\mathtt{methodC}$: \texttt{[callD]} applying this strategy.
\end{itemize}

In \name, loops (e.g., for, while) are viewed as paths that are repeatedly traversed in a tail-recursive way and are cycle-free. This crucial feature allows our approach to enumerate all possible execution paths. Nodes within the loop can occur multiple times in a row to mimic the actual execution sequences.
When acquiring LogEPs, we mark the nodes (i.e., first and last node) in a loop with a special sign.
For instance, LogEP for methodA will be marked as: \texttt{[Log@1$_{start}$, callB$_{end}$]}, and \texttt{[Log@1$_{start}$, callC$_{end}$]}.



Although LogEPs provide all possible executable paths, there exist some paths that are not executable under any input values. To avoid such infeasible paths, we conduct intra-procedural constraint analysis following previous studies~\cite{yuan2012improving}. In specific, we gather all constraints for each LogEP and filter out any LogEPs that contain unsatisfiable constraints.
This process makes the generated LogEPs more realistic.
Using the scalable traversal strategy, \name~obtains all LogEPs in less than 1.5 hours in an HDFS system that includes 3,749 methods (pruned) and 2,535 logging statements.






\subsection{PHASE3: Log Path Walking}
This phase is devised with the goal of generating labeled log sequences to simulate the actual application execution by chaining LogEPs from each method.
To achieve accurate labels while saving annotation effort, \name~uses a \textit{seed-propagation strategy} that involves experts identifying a set of anomaly LogEPs as seeds, followed by automatically propagating the labels of these anomaly LogEPs to all acquired LogEPs.
Labeled log sequences are eventually generated by a succession of random LogEP selection step which walks over the invoked methods. 
The choice of LogEPs can be controlled by setting specified data size, execution entrance and anomaly rate, allowing a flexible dataset that simulates execution logs in multiple scenarios.


\subsubsection{Seed Anomaly LogEP Annotation}

Since identifying anomalous logs is challenging and mostly relies on domain expert knowledge, we adopt the human-annotation process similar to existing log-based anomaly detection datasets~\cite{oliner2007supercomputers, lin2016log, xu2009detecting}. However, \name~differs from existing datasets in that, it can automatically propagate labels from a set of anomaly unit to the sequence-level, significantly improving the annotation efficiency.

We use LogEP as an anomaly unit for expert labeling as it is a small execution path reflecting the system behavior in a period.
The output of this step is a set of seed anomaly LogEPs that \textit{must} produce anomalies.
To obtain anomaly LogEPs, we design the annotation process with \textit{alerting statement annotation} and \textit{anomaly LogEP identification}.
The alerting statement annotation is designed to filter out a large number of normal logging statements (e.g., \texttt{Log@1}). To do so, we present all logging statements, their corresponding code snippets as the execution context, as well as the nearby comments inserted by developers, for annotators to decide whether they are alerting logging statements for an anomaly. Taking the HDFS system as an example, we present the example alerting statements and their corresponding potential anomalies in Table~\ref{tab:annotate-anomaly}.
However, identifying alerting statements is
not enough for profiling system activities.
Hence, anomaly LogEP identification aims to further determine whether LogEP will certainly lead to anomalies. To recognize anomaly LogEPs, we ask annotators to manually check each LogEP that contains alerting statements. The identified anomaly LogEP is considered anomaly seeds for propagation in the next step.

\begin{table}
\footnotesize
\centering
\caption{Alerting logging statements examples and their potential anomaly types in HDFS system.}
\begin{tabular}{c|l}
\toprule
    Anomaly Type & Alerting Logging Statements Examples\\
\midrule
\multirow{2}{*}{Version mismatch} & Layout version on remote node does not match this\\
 &  node's layout version.\\
\midrule
\multirow{2}{*}{Disk/Storage error} & Unable to get json from Item.\\
&  Unexpected health check result for volume $<*>$.\\
\midrule
\multirow{2}{*}{Dependency error} & Unresolved dependency mapping for host {}. \\
 & Continuing with an empty dependency list\\
\bottomrule
\end{tabular}
\label{tab:annotate-anomaly}
\end{table}

\subsubsection{Anomaly Label Propagation}

To generate an effective anomaly detection dataset, it is important to have labels at the sequence-level even though the annotations are done at the LogEP-level to save human effort.
To this end, \name~uses a strategy called seed propagation to propagate the labels of seed anomaly LogEPs to other LogEPs, with the goal of figuring out whether a LogEP is infected by the anomaly label.
The main idea is that: If a LogEP in one method contains an anomaly (e.g., $\mathtt{methodD}$), then all other LogEP invokes this method (e.g., \texttt{[callD]}) \textit{may} also induce an anomaly.

The propagation starts from the seed anomaly LogEPs marked \textit{infected}. The propagation is done recursively by checking whether a LogEP contains other infected LogEPs brought from invocations.
After the propagation, infected LogEPs are \textit{likely to}, but will not necessarily cause an anomalyz, whereas others \textit{must} be anomaly-free.



\subsubsection{Generating Log Sequences}
Given the LogEPs, \name~eventually generates each actual log sequence by selecting and chaining the LogEPs in a top-down approach. 
The top-down random walking process works as follows: It starts at an entrance method and walks along invocations, with one LogEP being randomly chosen at each walking step. If an invocation exists in the current chosen LogEP, \name~walks to the callee method and then chooses a LogEP in the callee method. The logging statements in all chosen LogEPs are chained according to the invocation relationships to form the log sequence.

As one log sequence reflects the execution path of a single thread, \name~generates a labeled sequence at each time with two walking strategies and combines the sequences to form the datasets for anomaly detection:
\begin{itemize}
    \item To generate an \textit{anomaly log sequence}, we always select the infected LogEP in every step until we have selected an anomaly LogEP that may contain alerting logging statements. 
    \item To generate a \textit{normal log sequence}, we randomly select a LogEP of each method but take a step back when selecting an anomaly LogEP, and re-choose another LogEP. 
\end{itemize}

During the walking, invocations or logging statements within the loop may occur successively more than once. 
The log sequence generation process with random path selection enables the flexibility of datasets. It can be decorated with a set of hyper-parameters to generate more controllable log sequences. For example, the component indicator (CI) controls the starting point to simulate the execution path in a specific component (e.g., storage component), and the anomaly rate $AR$ controls the anomaly ratio ($\frac{\#Anomaly\_Sequence}{\#All\_Sequence}$) in the generated dataset. 

\section{Implementation}\label{sec:implementation}
\subsection{Experiment Environment}
\name~has been implemented by 5,182 lines of Java code with Soot~\cite{vallee2010soot}, a Java bytecode optimization and analysis framework. We run all experiments on Ubuntu 18.04. The experiments are carried out on a machine  with an Intel(R) Xeon(R) Platinum 8255C CPU (@2.50GHz) with 128GB RAM. We set the $AR$ to be 3\% and $CI$ to be all possible paths to ensure the coverage, unless otherwise specified.

\subsection{Annotation}
To ensure the correctness of annotation, we invite three Ph.D. students who have at least two-year experience in distributed system research and development, two of whom annotate individually, and the other one works as an adjudicator to discuss the disagreements with annotators. They are all allowed to access the Internet for searching answers.
The agreement score between annotators measured by Cohen's kappa~\cite{banerjee1999beyond} before adjudication is 0.841 and 0.834 in alerting statement annotation and anomaly LogEP identification, respectively. All annotators reach a consensus on the labels after discussing them with the adjudicator.
\section{Experiments}\label{sec:exp}

We evaluate \name~using four research questions:

\noindent \textbf{RQ1:} How comprehensive are the datasets generated by \name?\\
\noindent \textbf{RQ2:} Is \name~scalable for real-world applications?\\
\noindent \textbf{RQ3:} How flexible are the datasets compared with passively-collected datasets?\\
\noindent \textbf{RQ4:} Can AutoLog benefit anomaly detection problems?

\begin{table*}[tbp]
\small
\centering

\caption{The comparison of datasets for comprehensiveness. $\mathcal{D}$-Coverage is reported for the systems with publicly log datasets.}
\begin{tabular}{c||c|c|c|c|c}
\toprule
System & Dataset & \# Log Event  & Logging Coverage & $\mathcal{D}$-Coverage & Increment ($\uparrow$)  \\ \midrule
\multirow{2}{*}{Hadoop} & $\mathcal{D}$-Hadoop   & 242 & 242/3426 (7.1\%) & \multirow{2}{*}{219/242 (90.5\%) } &  \multirow{2}{*}{\texttt{12x}}\\ 
& \name-Hadoop & 2879& 2879/3426 (84.0\%) &  & \\ \midrule
\multirow{2}{*}{HDFS} & $\mathcal{D}$-HDFS   & 30 & 30/1700 (1.8\%) & \multirow{2}{*}{27/30 (90.0\%)} & \multirow{2}{*}{\texttt{58x}}\\ 
& \name-HDFS & 1367 & 1367/1700 (80.4\%) &  & \\ \midrule
\multirow{2}{*}{Zookeeper} & $\mathcal{D}$-Zookeeper & 77 & 77/758 (10.2\%) & \multirow{2}{*}{77/77 (100\%)} & \multirow{2}{*}{\texttt{9x}} \\ 
& \name-Zookeeper & 740& 740/758 (97.6\%) &  & \\ \midrule
Apache Storm & \name-Apache Storm &1754 & 1754/1887 (93.0\%)& - & - \\
Flink & \name-Flink & 1574& 1574/1711 (92.0\%) & - & - \\
Kafka & \name-Kafka & 847 & 847/1002 (84.5\%) & - & - \\ \bottomrule
\end{tabular}
\label{tab:exp-comprehensiveness}
\end{table*}

\subsection{Experimental Settings}

\subsubsection{Existing Datasets}
To verify the effectiveness of \name, we choose three widely-used publicly available datasets collected from Java applications. We use $\mathcal{D}$-\textit{sys} and \name-\textit{sys} to denote the baseline datasets collected from system \textit{sys} and collected by~\name, respectively. Details of the datasets are illustrated as follows:
\begin{itemize}
    \item $\mathcal{D}$-Hadoop~\cite{lin2016log}. Hadoop is an open-source framework designed to store and process large-scale data efficiently. The dataset is collected via running two standard applications and injecting three types of failures for anomaly detection.
    \item $\mathcal{D}$-HDFS~\cite{xu2009detecting}. HDFS is a distributed file system for large-data storage, enabling high-throughput access to data. $\mathcal{D}$-HDFS is collected from a private cloud environment executing benchmark workloads with labeled anomalies.
    \item $\mathcal{D}$-Zookeeper~\cite{he2020loghub}. Zookeeper provides a centralized service to manage a large set of hosts (e.g., synchronization, configuration information management). $\mathcal{D}$-Zookeeper is collected in a lab environment for log analysis \textit{without labelling}.
\end{itemize}

Since this paper presents the first methodology that actively generates log datasets without deploying and running the system, we compare \name~with all existing Java-based log datasets in our research questions.

\subsubsection{Evaluation Subjects}
Apart from three projects associated with existing log datasets, we extensively evaluate the effectiveness of \name~on the most popular 50 projects from the Maven repository~\cite{Maven}, with more than 10,000 usage times for each. 
The selected projects include, but are not limited to, distributed streaming platforms (e.g., Kafka), core Java packages (e.g., Apache HttpClient), and unit testing frameworks (e.g., JUnit). 
Among them, we present the detailed result of three widely-studied distributed system projects below and report statistical results for other projects.
\begin{itemize}
    \item Apache Storm~\cite{storm}. It is a distributed real-time computation system, allowing large-scale data processing and high-velocity data streams.
    \item Flink~\cite{flink}. Flink is a stream processing engine that can handle scale system events with low latency. 
    \item Kafka~\cite{kafka}. Kafka is a distributed event storage and stream-processing platform applied in thousands of companies.
\end{itemize}

\subsubsection{Metrics}
\begin{itemize}
    \item \textit{Coverage}. Motivated by software testing studies~\cite{park2012carfast, chen2018automated}, we evaluate the comprehensiveness of \textit{logging coverage}, which measures the percentage of the discovered log events to the total log events designed in the application ($\frac{\#Log\_Event}{\#Total\_Log\_Event}$). We also use $\mathcal{D}$-Coverage to evaluate the ratio of log events in existing datasets that are covered by \name~($\frac{\#Log\_Event\_Covered\_by\_AutoLog}{\#Log\_Event\_in\_Existing\_Dataset}$).
    \item \textit{Execution Time}. To validate the scalability of \name, we report the program execution time for each system, which includes the time for code analysis and data generation in a single machine. 
\end{itemize}


\begin{figure}[tb]
    \centering
	\includegraphics[width=0.95\linewidth]{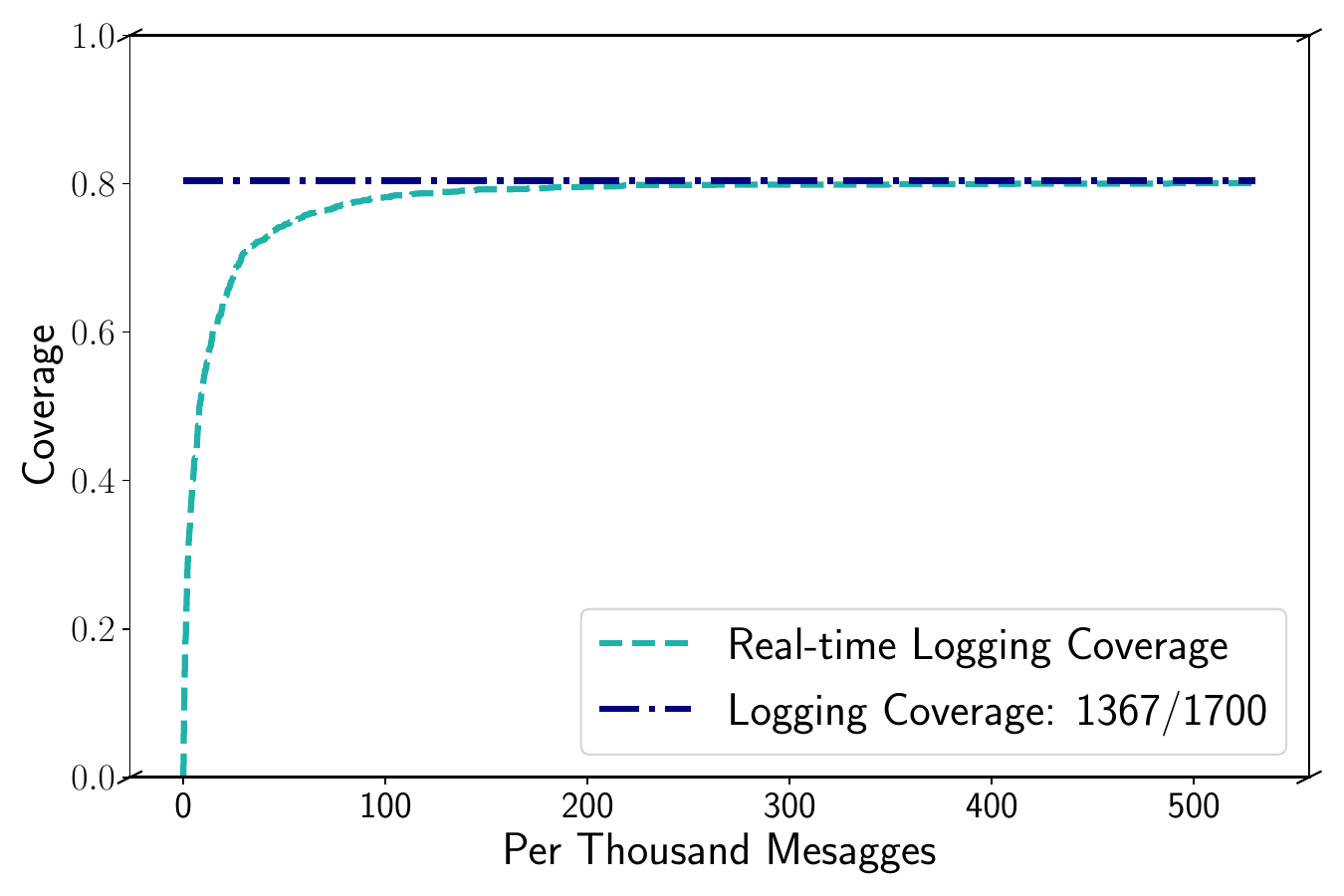}
	\vspace{-0.1in}
	\caption{The number of generated log messages and their corresponding real-time logging coverage.} 
	\label{fig:comprehensive-running}
\end{figure}



\subsection{RQ1: How Comprehensive are the Datasets Generated by \name?}

We evaluate the comprehensiveness of the dataset generated by \name~regarding the number of log events, logging coverage, and $\mathcal{D}$-Coverage, illustrated in Table~\ref{tab:exp-comprehensiveness}.

\begin{table*}
\small
\centering

\caption{The comparison of existing datasets for scalability.}
\begin{tabular}{c||c|c|c|c|c}
\toprule
System & Dataset  & \# Message  & Execution Time & \# Messages/min (speed) & Acceleration ($\uparrow$)  \\ \midrule
\multirow{2}{*}{Hadoop} & $\mathcal{D}$-Hadoop     & 394,308  & NA & NA & \multirow{2}{*}{\texttt{-}} \\ 
& \name-Hadoop & 392,427 & 3.41 hours & 1,918 & \\ \midrule
\multirow{2}{*}{HDFS} & $\mathcal{D}$-HDFS     &11,175,629  & 38.7 hours$^\dagger$ & 4,813& \multirow{2}{*}{\texttt{15x}}\\ 
& \name-HDFS & 11,376,233& 2.62 hours& 72,367& \\ \midrule
\multirow{2}{*}{Zookeeper} & $\mathcal{D}$-Zookeeper     & 207,820 & 26.7 days$^\dagger$ & 6 & \multirow{2}{*}{\texttt{2072x}} \\ 
& \name-Zookeeper &211,425 & 17 mins & 12,436&  \\ \midrule
Apache Storm & \name-Apache Storm & 1,001,245& 1.28 hours & 13,037 & - \\
Flink & \name-Flink & 1,003,416 & 1.21 hours & 13,821 & - \\
Kafka & \name-Kafka & 1,002,629 & 39 mins & 25,708 & - \\ \bottomrule
\multicolumn{5}{l}{$^\dagger$ This is the collection time from its original paper. NA means the authors do not report the collection time.}\\
\multicolumn{5}{l}{``-'' means the acceleration cannot be computed due to the unknown collection time of existing datasets.}
\end{tabular}
\label{tab:exp-scalability}
\end{table*}

Compared to the existing log datasets, \name~effectively enhances the comprehensiveness of log events. 
Firstly, \name~covers an average of 87.38\% logging statements over six systems, demonstrating its ability to capture logging statements designed in code. The number of log events generated by \name~is \texttt{12x, 58x,} and \texttt{9x} is more than existing datasets collected from Hadoop, HDFS, and Zookeeper, respectively. 
Regarding the missing parts, we analyze them as follows.
They mainly come from the restrictive call graph construction phase~\cite{yuan2012improving}, the limitation of logging statement restoring~\cite{utture2022striking} across different methods, and unreachable execution paths that we have eliminated. Taking the restoring process as an example,
if the constant string $\mathtt{msg}$ is defined outside $\mathrm{methodD}$ in Listing~\ref{lst:emp-case}, \name~has difficulty restoring the \texttt{Log@4}.
Secondly, the experiment results illustrate that \name~covers most of the log events in the existing dataset, achieving 219/242, 27/30, and 77/77 for Hadoop, HDFS, and Zookeeper systems, respectively. The missing log events mainly come from optional components of complicated systems and different deployment settings on varied platforms.

Logging coverage implies the upper limit of the discovered log events in \name; ideally, generating log sequences for enough time will eventually reach that coverage. Fig.~\ref{fig:comprehensive-running} shows the relationship between the number of real-time generated log messages and its corresponding logging coverage in HDFS system (denoted as real-time logging coverage).
The results indicate that \name~achieves its logging coverage after generating approximately 350,000 messages.


\begin{figure}[tb]
    \centering
	\subfigure[Logging coverage histogram]{
		\includegraphics[width=0.7\linewidth]{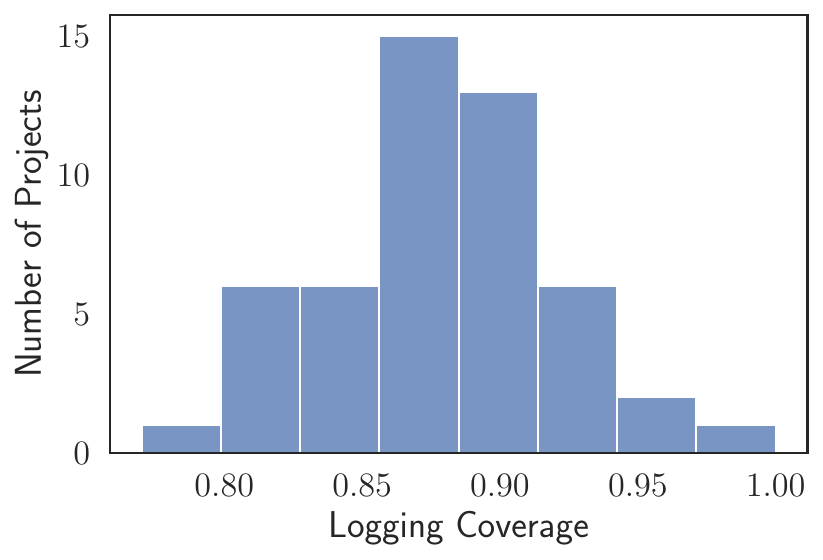}
	}       
	\subfigure[\# Log event histogram]{
		{\includegraphics[width=0.7\linewidth]{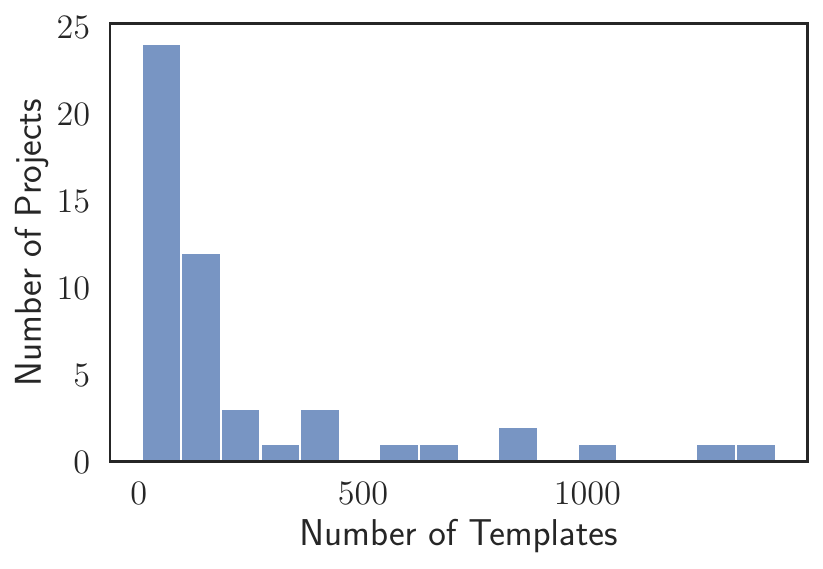}}}

	\caption{The histogram of logging coverage and the number of log events over 50 popular projects.} 
	\vspace{-0.1in}
	\label{fig:RQ2-large}
\end{figure}

Besides, we display the logging coverage (left) and the number of generated log events (right) histogram over 50 popular Java projects in Fig.~\ref{fig:RQ2-large}. The results demonstrate that \name~achieves an average logging coverage of 87.77\%, which is superior to existing log datasets. 
Additionally, the number of log events in different projects ranges from hundreds to thousands, indicating that existing log datasets with limited log events are inadequate.



\begin{tcolorbox}[boxsep=1pt,left=2pt,right=2pt,top=3pt,bottom=2pt,width=\linewidth,colback=white!90!black,boxrule=0pt, colbacktitle=white!30!viol,toptitle=2pt,bottomtitle=1pt,opacitybacktitle=0.4]
\name~shows a significant improvement (\texttt{9x-58x}) on the number of log events and 
can effectively cover \textit{87.77\%} logging statements on average over studied projects.
\end{tcolorbox}

\subsection{RQ2: Is \name~Scalable for Real-world Applications?}

We evaluate the scalability of our log generation methodology, which measures the time required for generating data. To compare the scalability, we generated the same amount of data in a single machine as existing public datasets and compared the data collection time. 

\begin{figure}[tbp]
     \centering
     \includegraphics[width=\columnwidth]{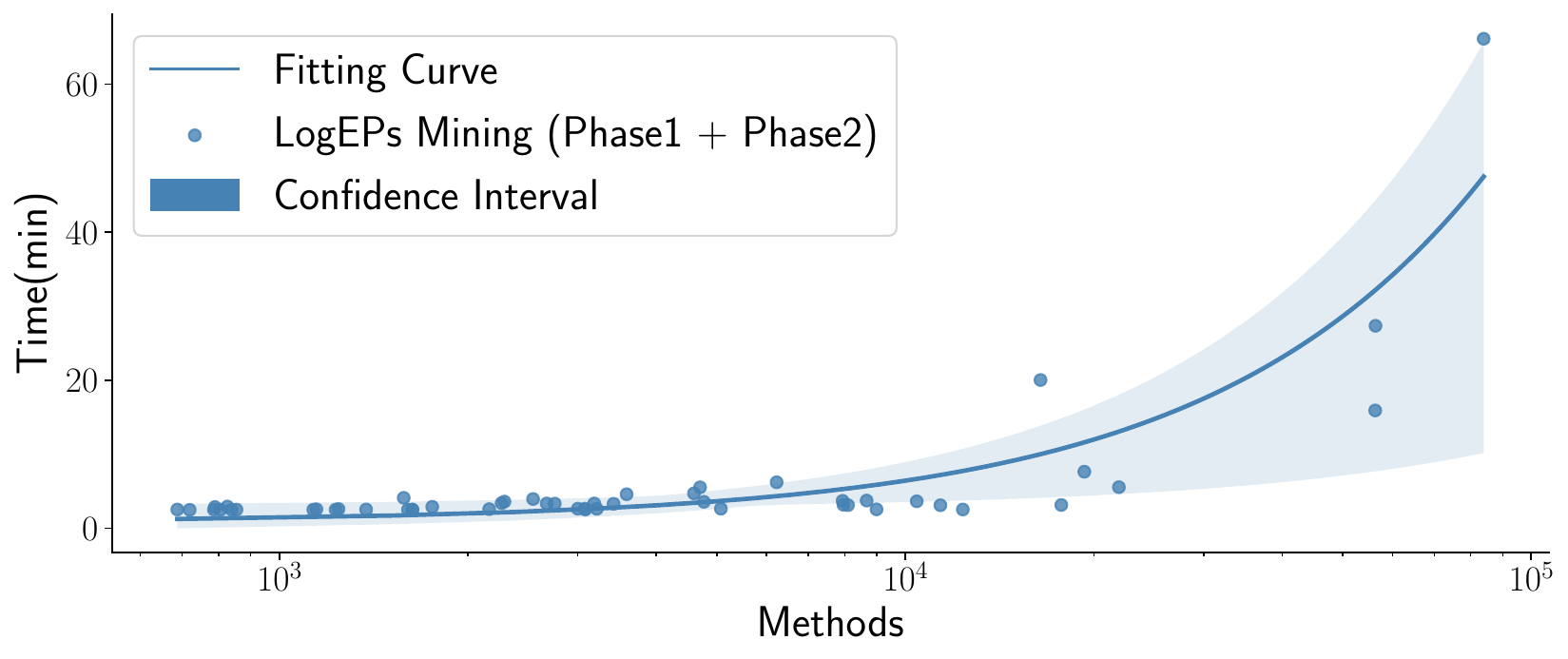}
     \vspace{-0.2in}
     \caption{Time cost for acquiring LogEPs and the number of analyzed methods over 50 popular projects.}
     \vspace{-0.2in}
     \label{fig:exp-scalability-trend}
 \end{figure}

The result in Table~\ref{tab:exp-scalability} indicates that \name~is efficient in analyzing and generating log sequences from real-world applications. It can produce messages at high speed, with a range of 1,918 to 72,367 messages per minute.
Moreover, compared with the HDFS dataset that took 38.7 hours to collect, \name~can generate the same amount of data within 2.62 hours (15 times faster).
This scalability is attributed to the call graph pruning and intra-method LogEP derivation steps, which decrease the number of methods for analysis and solve the path explosion problem.
Moreover, because \name~is built on source code, deploying it into a new system is effortless, unlike the long time required for configuring and rerunning applications in existing passively-collection methods.
Additionally, we investigate the time spent for execution path finding (Phase1 and Phase2) as LogEPs can be reusable to generate log sequences multiple times. In particular, we apply \name~on 50 Java projects and measure the execution time, ranging from 687 to 83,969 methods in a project (as shown in Fig.~\ref{fig:exp-scalability-trend}). Our results demonstrate that \name~can analyze most projects within one hour, indicating its practicality and efficiency.

Although we did not include the annotation time in Table~\ref{tab:exp-scalability}, \name~has an efficient seed-propagation labeling strategy that requires less time than labeling each log sequence in passively-collected datasets. For instance, independent annotators spent an average of 2 hours on alerting statement annotation and 4 hours on anomaly LogEP identification for 1,367 log events in HDFS.


\begin{tcolorbox}[boxsep=1pt,left=2pt,right=2pt,top=3pt,bottom=2pt,width=\linewidth,colback=white!90!black,boxrule=0pt, colbacktitle=white!30!viol,toptitle=2pt,bottomtitle=1pt,opacitybacktitle=0.4]
\name~considerably shortens the dataset generation time (\texttt{15x}) compared with existing datasets. The promising path finding
time further manifests its scalability for real-world applications.
\end{tcolorbox}

\subsection{RQ3: How Flexible are the Datasets Compared with Passively-collected Datasets?}\label{sec:RQ3}

We assess the flexibility of~\name~and its benefits for log-based anomaly detection. 
In Table~\ref{tab:exp-flexibility}, we compare \name~with existing datasets regarding the flexibility over \textit{data size}, \textit{component-indicator}, and \textit{anomaly rate}.

\begin{table}[tbp]
\footnotesize
\centering
\caption{The comparison of log datasets for flexibility.}
\begin{tabular}{c||c|c|c}
\toprule
Aspects & $\mathcal{D}$-HDFS & $\mathcal{D}$-Hadoop & \name\\
\midrule
\textbf{Data Size} & Partial & Partial & \textbf{Complete}\\
\midrule
\textbf{Component} & \multirow{2}{*}{Partial} & \multirow{2}{*}{Partial} & \multirow{2}{*}{\textbf{Complete}}\\
\textbf{Indicator} & & & \\
\midrule
\textbf{Anomaly Rate} & Deterministic & Deterministic & \textbf{Non-deterministic}\\
\bottomrule
\end{tabular}
\label{tab:exp-flexibility}
     \vspace{-0.1in}
\end{table}

Different systems require varying amounts of data for algorithm development. For example, a large amount of data can be exploited to build algorithms for distributed cloud systems whereas naive systems only preserve a small amount of data to study. \name~is capable of generating datasets of \textit{any} size, whereas $\mathcal{D}$-HDFS and $\mathcal{D}$-Hadoop have partial size restrictions once they are released.


In addition, engineers may need to focus on specific components' functionality during software development and maintenance, which requires component-specific logs for anomaly detection.
While the existing datasets provide limited component information, \name~is able to produce sequences of component-indicator logs by starting to walk from constrained entry nodes during generation.

Moreover, different systems have different fault-tolerance abilities, which affect the potential anomaly rates in collected logs. Existing datasets (e.g., $\mathcal{D}$-HDFS) have fixed numbers of anomaly sequences, requiring researchers to filter out some anomaly sequences or significantly lower the number of normal sequences to increase the anomaly rate in a deterministic way~\cite{le2022log}. However, \name~can produce a huge amount of various sequences iteratively, introducing the unseen patterns and increasing the flexibility for anomaly detection.



\begin{tcolorbox}[boxsep=1pt,left=2pt,right=2pt,top=3pt,bottom=2pt,width=\linewidth,colback=white!90!black,boxrule=0pt, colbacktitle=white!30!viol,toptitle=2pt,bottomtitle=1pt,opacitybacktitle=0.4]
\name~effectively generates datasets with more flexibility of utilization (i.e., data size, component, anomaly rate) than existing datasets, allowing imitating a wider range of sophisticated application scenarios.
\end{tcolorbox}

\subsection{RQ4: Can AutoLog Benefit Anomaly Detection Problems?}\label{sec:discussion}


This RQ explores how \name~helps to resolve log-based anomaly detection.
In particular, we train state-of-the-art (SOTA) detectors on \name~and evaluate their performance. 
This section takes HDFS\footnote{Other widely-used anomaly detection datasets (e.g., BGL) are not generated from open-sourced software.} to evaluate models on real-machine-generated logs $\mathcal{D}$-HDFS (denoted as $\mathcal{D}$) and \name-HDFS (denoted as \name) from the same software version, followed by a performance discussion.


\subsubsection{Settings} We select the following representative log anomaly detection models for evaluation:
(1) LogRobust~\cite{zhang2019robust}, which leverages a bi-directional LSTM network (bi-LSTM) with an attention mechanism to learn the importance of each log for tackling unstable log patterns. (2) CNN~\cite{lu2018detecting}, which transforms the log sequence to a trainable matrix, then applies a Convolutional Neural Network (CNN) for log-based anomaly detection. (3) Transformer~\cite{nedelkoski2020self}, which applies a Transformer encoder to learn context information to distinguish the anomaly logs from normal logs.
When using \name, we set the anomaly rate (3\%), and data split ratio (train: test = 8: 2) to be the same as the original $\mathcal{D}$. 

Following previous research work~\cite{zhang2019robust,le2021log}, we adopt Precision (P), Recall (R), and F1 to evaluate the performance. \textit{Precision} is calculated by the percentage of log sequences correctly identified with anomalies over all sequences that are recognized with anomalies. \textit{Recall} is calculated by the percentage of log sequences correctly recognized with anomalies over the actual anomaly sequences. \textit{F1 Score} (F1) is the harmonic mean between Precision and Recall.

\subsubsection{Results}



\begin{table}[tbp]
\footnotesize
    \centering
    \caption{Comparison of the anomaly detection models over two datasets $\mathcal{D}$ and \name.}
        \begin{tabular}{cl||c|c}
    \toprule
   & Train set & $\mathcal{D}$ & \name \\
   \midrule
   Test set & Approach & P \quad R  \quad F1 & P \quad R \quad F1  \\
    \midrule
       & Transformer & 0.889  0.904  0.896 & 0.892 0.996 \textbf{0.941}  \\
      $\mathcal{D}$ & CNN & 0.936  0.995  0.965 & 0.959 0.997 \textbf{0.978} \\
       & LogRobust & 0.942 0.994 \textbf{0.967} & 0.947 0.988 \textbf{0.967} \\
    \midrule
       &  Transformer & -$^\dagger$  & 0.723  0.755  0.739  \\
     \name & CNN & -$^\dagger$  & 0.697  0.790  0.741 \\
       & LogRobust & -$^\dagger$  & 0.673  0.875  0.761  \\
    \bottomrule
    \multicolumn{4}{l}{$^\dagger$ The large amount of unseen events causes considerably poor performance.}
    \end{tabular}

    \label{tab:discussion-compare}
\end{table}

Table~\ref{tab:discussion-compare} presents the performance of different anomaly detection models on two datasets. 
First, we evaluate models on the real-world dataset $\mathcal{D}$ (test set). We observe that models trained with \name~perform \textit{consistently better} (1.93\% of F1 on average) than trained with $\mathcal{D}$ (train set). CNN can achieve an F1 score of 0.978 after training with \name, even surpassing the SOTA performance by 1.1\%.
We analyze the reason for performance improvement after training on \name as follows. \name~generates more comprehensive log events by imitating a more varied system behavior. Once an anomaly detector has seen such diverse log events in its training phase, it can effectively detect the anomalies in production environments.
The result demonstrates that \name~is effective in generating realistic log sequences, which can help models learn more varied system behavior and improve anomaly detection accuracy.

Second, we observe that state-of-the-art approaches perform well in $\mathcal{D}$ dataset but can only reach an F1 score of 0.761 in \name. We attribute the performance gap to the more comprehensive log events (e.g., 1367 rather than 30) and diverse sequential log patterns generated by \name.
This highlights the need for further research on 
log semantic encoding and system behavior profiling to address the challenges posed by complex log data in real-world deployments of anomaly detection models.


\begin{tcolorbox}[boxsep=1pt,left=2pt,right=2pt,top=3pt,bottom=2pt,width=\linewidth,colback=white!90!black,boxrule=0pt, colbacktitle=white!30!black,toptitle=2pt,bottomtitle=1pt,opacitybacktitle=0.4]
\name~benefits anomaly detection detectors by providing the training resource that allows existing models to improve (\textit{1.93\%} on average) their performance consistentl.
It is effective in generating more sophisticated and practical log data than any existing Java log datasets.
It can serve as a benchmark data generator and training resource for developing and validating promising anomaly detection models. 
\end{tcolorbox}



\section{Case Study}

\begin{figure}[tbp]
     \centering
     \includegraphics[width=\columnwidth]{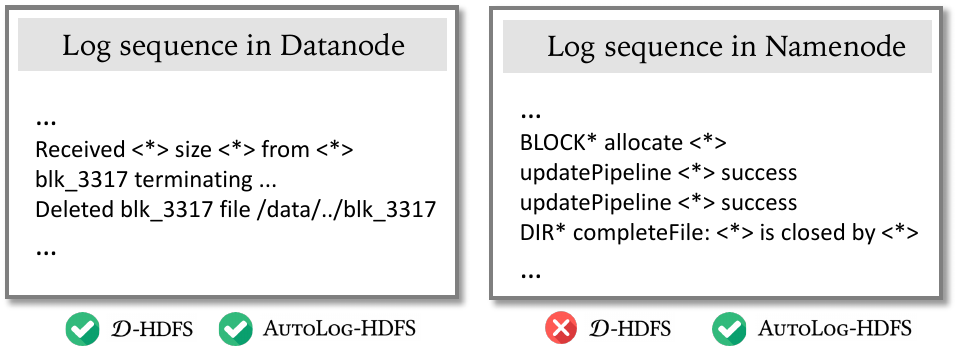}
          \vspace{-0.1in}
     \caption{User-reported log event sequences from HDFS.}
     \label{fig:case-study}
 \end{figure}

Fig.~\ref{fig:case-study} exhibits a case from an HDFS user who tries to delete blocks and encounters data loss issue\footnote{https://issues.apache.org/jira/projects/HDFS/issues/HDFS-16829}. The user reports the log sequences from Datanode (left) and Namenode (right). 
After matching this realistic log sequence in $\mathcal{D}$-HDFS and \name-HDFS, we find that (1) both $\mathcal{D}$-HDFS and \name-HDFS contain the sequence from Datanode, (2) only \name-HDFS can cover the sequence from    Namenode.
In \name, the acquired execution paths via program analysis simulate how logging statements are executed (recorded) in software to provide realistic log sequences in block deletion.
In addition, the case illustrates the comprehensiveness of logs produced by \name. Although it is impractical to configure and enumerate all system activities in software for log collection, \name~addresses the issue by actively analyzing all possible execution paths.

\section{Threats to Validity}\label{sec:threats}

We have identified the following four major threats to validity.
(1) \textit{Unpredictable exception handlers}:
A few run-time behaviors, such as exception catching, cannot be thoroughly analyzed without actually executing the program. A logging statement can be severed from the execution path by the exception handler. 
While exceptions can reflect anomalies, these anomalies are relatively ``easy'' to be detected (e.g., keyword search). Existing log-based anomaly detection mainly focuses on the remaining ``difficult'' anomalies.
To mitigate this threat, we can simulate run-time exception-catching behaviors by randomly selecting try-catch statements and setting interruptions in the try body.

(2) \textit{Potential multi-thread intervention}:
\name~analyzes execution paths and generates log sequences in a single thread. Thus, the intervention and communication between multiple threads may be neglected. However, we notice that modern anomaly detection models~\cite{zhang2019robust,lu2018detecting,nedelkoski2020self} are developed for single-thread analysis, which starts with separating different threads based on their thread IDs.
In this regard, the lack of multi-thread log sequences will not hamper the evaluation of existing anomaly detection models.
In the future, we plan to extend \name~to support multi-thread settings.

(3) \textit{Unresolved parameters}:
\name~uses a dummy token to replace unresolved parameters (e.g., addresses, digits, ID character strings) in the logging statements, which is supposed to carry system behavior information. 
However, previous studies~\cite{wang2021groot, chen2021experience} reveal that log event (without run-time parameters) sequences are sufficient to capture system activities, and parameters may hinder deep learning-based detectors~\cite{zhang2019robust}.
Even though the unresolved parameters are specified with certain values, they will be wiped out before feeding into anomaly detection models.
In any case, we regard parameters resolving as an important future direction for their utilization in other log analytics (e.g., log parser).

(4) \textit{Imprecise call graph}: Generating precise call graphs has been known as an open-challenging question in static analysis community for a long time~\cite{sadowski2018lessons, johnson2013don, utture2022striking}. Soot~\cite{vallee2010soot} uses Class Hierarchy Analysis (CHA)~\cite{dean1995optimization} to handle dynamic dispatch (e.g., finding the callee), which has a relatively low accuracy due to the polymorphism of Java. To mitigate the impact of the imprecise graph, we refined the calling relationship with context-insensitive pointer analysis that significantly improves the precision of the call graph.

\section{Related Work}\label{sec:relatedwork}

\subsection{Logging Statements Generation}

Recording the precise and concise logs in software becomes a critical problem for logging practice as developers inspect logs for software operation and maintenance.
Logging statements generation aims at suggesting developers in deciding \textit{what} to log, \textit{where} to inject a log and at which \textit{log level}.
A collection of studies are proposed to support developers in logging activities. For \textit{where-to-log}, \cite{li2020shall} extracted syntactic and semantic information from the code block level to suggest the logging locations via a deep learning-based approach.
For \textit{log level}, PADLA~\cite{mizouchi2019padla} is presented to dynamically adjust the log level of a running system guided by the online phase performance anomalies detection. Similarly, DeepLV~\cite{li2021deeplv} is a neural approach that automatically suggests the log level by incorporating the syntactic context (i.e., AST) and message features (i.e., log message) from code. Regarding \textit{what-to-log}, several studies suggest the variables to log via source-code analysis~\cite{yuan2012improving} and variable representation learning~\cite{liu2019variables}.
The recent study~\cite{mastropaolo2022using} applied a Text-To-Text-Transfer-Transformer (T5) model to generate complete logging statements and inject them in the correct code location. The latest study~\cite{li2023exploring}, inspired by the huge success of large language models (LLMs) in natural language understanding and code intelligence tasks, empirically analyzed how well LLMs in generating complete logging statements.
Instead of deciding on log statements in code, \name~stands at a completely different perspective that automatically generates log sequences from existing projects for the research and development community.

\subsection{Data Synthesis}

Data synthesis has been investigated in various domains and modalities, such as time series and traces. Statistical models~\cite{wang2021tsagen} and generative models~\cite{yoon2019time, esteban2017real} are two classical methodologies to synthesize time-series data.
For example, TSAGen~\cite{wang2021tsagen} is a univariate time-series synthesizer that yields Key Performance Indicator (KPI) data with various anomalies and controllable characteristics using a probability distribution. Moreover, a few studies have been performed to synthesize location traces to protect mobile users' privacy, including employing synthetic point injection in a trajectory of a user~\cite{shokri2011quantifying}, synthesizing differentially private trajectories by variable-length n-grams~\cite{chen2012differentially} or hierarchical reference systems~\cite{he2015dpt}. \name~differs from all above studies in the way that it generates data from the code level to simulate execution paths instead of synthesizing from existing data.

The only study related to synthesizing logs is LogRobust~\cite{zhang2019robust}, which enhances the existing dataset by randomly shuffling, adding, and removing words in collected logs. However, such a mutation-based synthesis strategy can hardly reflect the real logging statements and their sequential patterns that are not covered in data collection. 


\subsection{Log-based Anomaly Detection}

Automated log file analysis enables to detect anomalous system activities early and guides troubleshooting. The anomaly detection task requires the model to identify whether there exist anomalies in a short time of logs. 
Several machine learning-based models (e.g., decision tree~\cite{chen2004failure}, support vector machine~\cite{liang2007failure}) are built to solve the task. Recent studies demonstrate that the advanced deep learning-based models~\cite{huo2023evlog} achieve higher performance, such as LSTM~\cite{du2017deeplog, zhang2019robust}, auto-encoder~\cite{farzad2020unsupervised}, and Transformer~\cite{nedelkoski2020self, le2021log}.

Although plenty of efforts have been devoted to intelligent log anomaly detection, log files are seldom made available to the public, making it difficult for researchers to validate new algorithms. The most widely-used log analysis dataset, LogHub~\cite{he2020loghub}, only contains limited logging statement coverage and thus cannot reflect the modern complex systems.
\name, as a log dataset generation methodology, breaks this bottleneck by efficiently providing a comprehensive and controllable data source.
\section{Conclusion} \label{sec:conclusion}
Although log-based anomaly detection has been widely studied, only a few approaches have been successfully deployed in the real world because existing datasets suffer from comprehensiveness, scalability, and flexibility limitations.
To overcome these limitations, this paper presents \name, the first automated log generation methodology for anomaly detection, with three phases: logging statement probing, log-related execution path finding, and log path walking.
Extensive experiments demonstrate that \name~produces comprehensive coverage of log events in application with scalability. \name~is also equipped with hyper-parameters to generate log datasets in flexible data size, component indicator, and anomaly rate. 
With our replication package released, we believe \name~could be a starting point for active dataset generation in the log analysis field, which provides benchmarking data for building practical anomaly detection algorithms.

\section{Acknowledgment}

The work described in this paper was supported by the National Natural Science Foundation of China (No. 62202511), the Key-Area Research and Development Program of Guangdong Province (No. 2020B010165002), and the Key Program of Fundamental Research from Shenzhen Science and Technology Innovation Commission (No. JCYJ20200109113403826). It was also supported by the Research Grants Council of the Hong Kong Special Administrative Region, China (No. CUHK 14206921 of the General Research Fund). We thank all reviewers for their valuable comments.
\newpage
\balance
\bibliographystyle{IEEEtran}
\bibliography{autolog}

\end{document}